\documentclass[12pt, preprint]{aastex}

\usepackage{graphicx, xspace}

\newcommand{\etal}{{\it et al.}\xspace}

\newcommand{\einstein}{{\it Einstein}\xspace}
\newcommand{\rosat}{{\it ROSAT}\xspace}
\newcommand{\chandra}{{\it Chandra}\xspace}

\newcommand{\NH}{\mbox{$N_{\rm H}$}\xspace}        

\begin{document}

\title{The X-ray Halo of GX 13+1}
\author{Randall K. Smith, Richard J. Edgar}
\affil{Smithsonian Astrophysical Observatory, 60 Garden St.,
Cambridge, MA 02138}
\email{rsmith,redgar@cfa.harvard.edu}
\and
\author{Richard A. Shafer}
\affil{Code 685, Laboratory for Astronomy and Space Physics, NASA
Goddard Space Flight Center, Greenbelt, MD 20771}
\email{richard.shafer@gsfc.nasa.gov}

\begin{abstract}
We present observations of the X-ray halo around the LMXB GX~13+1 from
the \chandra X-ray telescope.  The halo is caused by scattering in
interstellar dust grains, and we use it to diagnose the line-of-sight
position, size distribution, and density of the grains.  Using the
intrinsic energy resolution of \chandra's ACIS CCDs and the recent
calibration observation of the \chandra point spread function (PSF),
we were able to extract the halo fraction as a function of energy and
off-axis angle.  We define a new quantity, the ``halo coefficient,''
or the total halo intensity relative to the source at 1 keV, and
measure it to be $1.5^{+0.5}_{-0.1}$\ for GX~13+1.  We find a
relationship between this value and the dust size, density, and
hydrogen column density along the line of sight to GX~13+1.  We also
conclude that our data does not agree with ``fluffy'' dust models that
earlier X-ray halo observations have supported, and that models
including an additional large dust grain population are not supported
by these data.
\end{abstract}

\section{Introduction}

X-ray halos are created by the small-angle scattering of X-rays
passing through dust grains in the interstellar medium.  The spectral
and spatial characteristics of X-ray halos are determined by the size,
line of sight distribution, and composition of dust grains, with a
bias towards larger dust grains which are the primary sites for X-ray
scattering.  These larger grains hold much of the mass that is in dust
grains and they are difficult to detect in other wavebands.  In this
paper, we will describe how observations of the halo around GX~13+1
can be used to place limits on dust grain models, specifically on the
population of larger grains.

GX~13+1, a low-mass X-ray binary (LMXB), is a bright highly-absorbed
X-ray source.  IR spectral observations by Bandyopadhyay \etal (1999)
show that the secondary star is likely a K5 giant, and they calculate
a distance of $7\pm1$\,kpc for the system.  GX~13+1 has been
classified as an atoll source (Hasinger \& van der Klis 1989),
although it belongs to the small subclass of atoll sources that are
persistently bright.  The X-ray halo around GX~13+1 was first
described by Catura (1983), who used \einstein HRI data and measured a
total halo flux equal to $0.18\pm0.02$\ of the source flux in the HRI
band (0.1-6 keV).  However, Catura noted that systematic uncertainties
in the observation probably lead to an underestimate of the halo
strength.  Subsequently, Mauche \& Gorenstein (1986) used \einstein
IPC observations of GX~13+1 to measure a halo fraction of
$0.17\pm0.01$ in the IPC band (0.1-6.4 keV).  In 1995, Predehl \&
Schmitt measured the X-ray halo of GX~13+1 using only 47 seconds of
observations from the ROSAT All-Sky Survey.  They found a much
brighter X-ray halo, $0.37\pm0.04$\ of the source flux in the PSPC
band (0.08-2.9 keV).  The discrepancy between the \einstein and \rosat
observations has not been discussed, but we will show it is due to the
different energy bands of the instruments.

Measurements of X-ray halos have been used to place limits on the dust
grain density (Mathis \etal 1995), as well as the dust size
distribution (Witt, Smith \& Dwek 2001).  The \chandra observations
provide the far higher angular and energy resolution than has been
available, so we can extract the halo fraction as a function of
off-axis angle and energy.  By comparing the observed halo fraction to
a selection of dust models, we put limits on the density of the dust
and the relative population of large grains.

\section{Theoretical Background}

The theory of X-ray halos was first discussed in an astrophysical
context by Overbeck (1965), and has since been refined by a number of
authors (Mauche \& Gorenstein 1986; Mathis \& Lee 1991; Smith \& Dwek
1998).  We briefly review the theory here.

The fundamental quantity is the differential scattering cross section
$d\sigma/d\Omega$, which can be calculated using either the exact Mie
solution or the Rayleigh-Gans (RG) approximation; see Smith \& Dwek
(1998) for a discussion.  The RG approximation is derived by assuming
each volume $dV$\ in the dust scatters X-rays via Rayleigh scattering,
and then integrating the result over the grain volume.  Analytically,
this gives
\begin{equation}
{{d\sigma(\theta_{\rm sca})}\over{d\Omega}} = 1.1\,\hbox{cm}^2\hbox{sr}^{-1}
\Big({{2Z}\over{M}}\Big)^2 \Big({{\rho}\over{3 \hbox{g\,cm}^{-3}}}
\Big)^2 a_{\mu\rm m}^6 \Big({{F(E)}\over{Z}}\Big)^2
\Phi^2(\theta_{\rm sca})
\label{eq:RG}
\end{equation}
where $a$\ is the grain radius, $Z$\ is the mean atomic charge, $M$\
the mean atomic weight (in amu), $\rho$\ the mass density, $E$\ the
X-ray energy in keV, $F(E)$ the atomic scattering factor (Henke 1981),
$\theta_{\rm sca}$\ the scattering angle, and
$\Phi^2(\theta_{\rm sca})$\ the scattering form factor (Mathis \&
Lee 1991).  For 
homogeneous spherical grains, the form factor is given by 
\begin{equation}
\Phi^2(\theta_{\rm sca}) = 3(\sin u - u \cos u)/u^3
\end{equation}
where $u = 4\pi a \sin(\theta_{\rm sca}/2)/\lambda \approx 2\pi a
\theta_{\rm sca} E/hc$.  

Smith \& Dwek (1998) showed that the RG approximation will
overestimate the total scattering if the energy of the X-rays (in keV)
is not substantially larger than the size of the dust grains (in
$\mu$m), and suggested 2 keV as a minimum energy for most ISM dust models.

By integrating the scattering cross section over the line of sight
geometry, the source spectrum, and the dust size distribution we get
(considering single scatterings only) the halo surface brightness at
angle $\theta$\ from the source:
\begin{equation}
I_{\rm sca}(\theta) = F_X N_H \int dE\,S(E) \int da\,n(a) \int_0^1 dx\,
{{f(x)}\over{(1-x)^2}} {{d\sigma}\over{d\Omega}}
\end{equation}
where $F_X$\ is the total source flux, \NH is the hydrogen column
density, $S(E)$\ is the (normalized) X-ray spectrum, and $n(a)da$\ is
the dust grain size distribution.  Here $f(x)$\ is the density of
hydrogen at distance $xD$\ from the observer divided by the line of
sight average density, where $D$\ is the distance to the source
(Mathis \& Lee 1991).

Of course if the column density is sufficiently large, individual
X-rays may be scattered multiple times.  Mathis \& Lee (1991) showed
that for $\tau_{\rm sca} > 1.3$, multiple scatterings dominate over
single scattering, tending to broaden the halo.  The scattering cross
section depends upon the X-ray energy and the dust model; Table 1 of
Mathis \& Lee (1991) shows that $\sigma_{\rm sca} = 9.03\times10^{-23}
E^{-2}_{\rm keV}$ for dust models such as Mathis, Rumpl \& Nordsieck
(1977; MRN) or Draine \& Lee (1984).  For \NH $\approx
2.9\times10^{22}$\,cm$^{-2}$\ (see \S\ref{subsec:spectral}), this
corresponds to $\tau_{\rm sca} = 2.6 E_{\rm keV}^{-2}$.  Above 2 keV,
therefore, multiple scattering should not be a significant effect.
Below 2 keV, we can expect that the RG approximation will somewhat
overestimate the total halo intensity and the single scattering
approximation will underestimate the radial extent of the halo.

\section{Observations and Analysis}

\begin{figure}
\includegraphics[totalheight=3in]{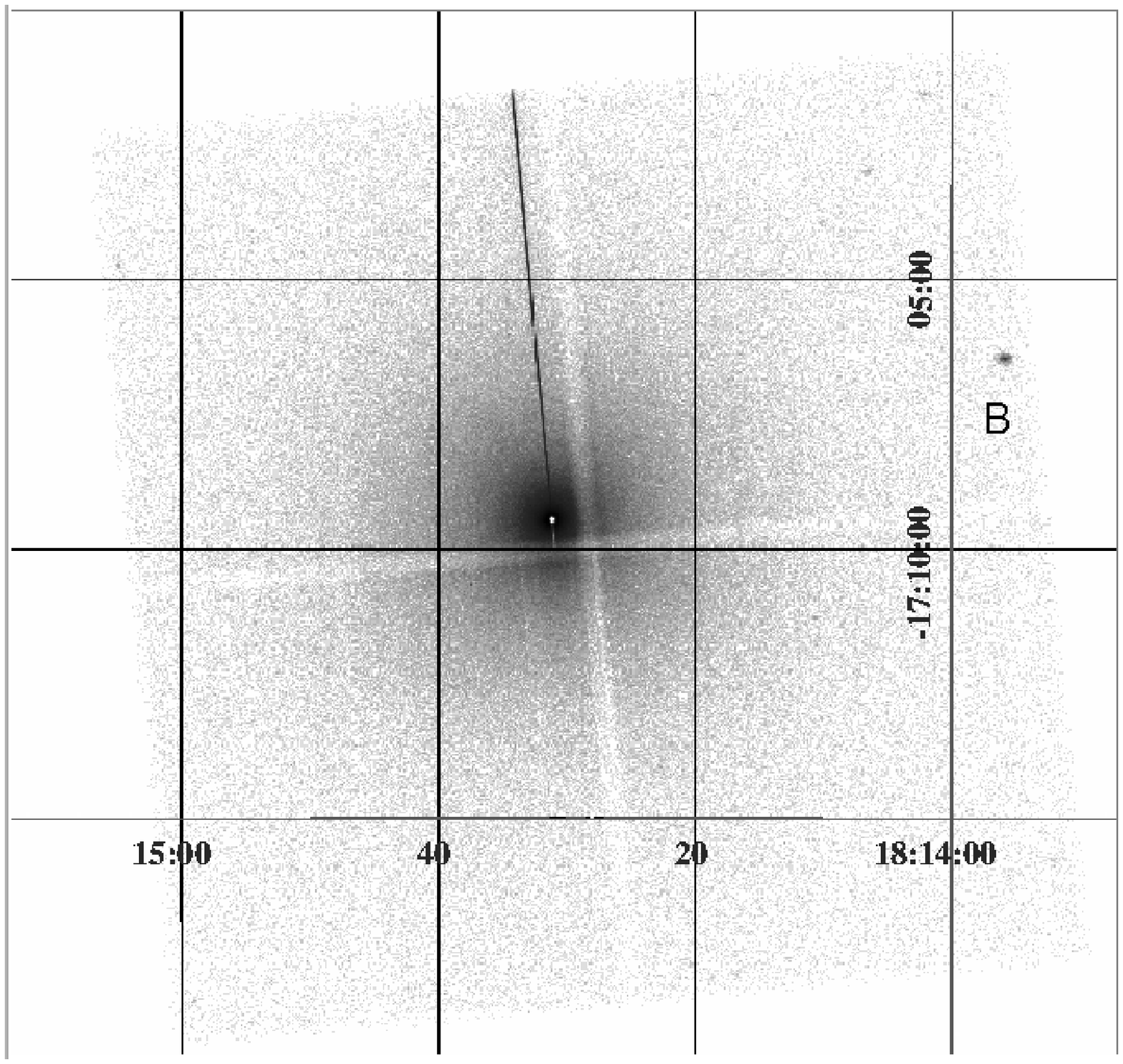}
\caption{\chandra ACIS-I image of GX~13+1.  The transfer streak can be
seen extending from the central source, and the serendipitous source
``B'' is marked.  Also notable are the chip gaps of the ACIS-I
array.\label{fig:image}}
\end{figure}

GX~13+1 was observed with the \chandra ACIS-I array (chips I0-3, S2,
S3) for 9.74 ksec on August 7th, 2000.  GX~13+1 was at the aimpoint,
and as can be seen in Figure~\ref{fig:image}, a bright halo was
observed although as expected the source itself suffered from severe
pile-up.  We processed the data using the \chandra data system
software version R4CU5UPD14.1, and used CIAO version 2.2 for our
analysis.

The ACIS-I CCDs were damaged early in the \chandra mission by
low-energy protons.  As a result the energy resolution decreased; at
1.5 keV, the FWHM on ACIS-I3 now varies from $\sim 100$\,eV to $\sim
150$\,eV, depending on the distance from the CCD readout.  Based on this resolution and
the quantity of data, we decided to bin the data into 100 eV and 200
eV bins (for energies less than and greater than 2 keV, respectively). 

\subsection{Spectral Analysis\label{subsec:spectral}}

The X-ray halo is directly proportional to the source spectrum, so any
uncertainties in the spectral analysis will be reflected in the halo
analysis.  Ideally, the spectral measurement would be high-resolution
and contemporaneous; for this observation, however, we have only the
ACIS-I data, which are heavily affected by pileup.  In CCD detectors
such as the ACIS-I, pileup occurs when two or more photons impact the
same or adjacent pixels within a single 3.2 second frame.  This can
mimic a single photon with energy equal to the sum of the photon
energies, or it can change the event ``grade'' from a good X-ray
detection (grades 0, 2, 3, 4, or 6) to a rejected likely cosmic-ray
event (grade 7) (Chandra Proposers' Observatory Guide, Rev. 4.0, p
106).  In GX~13+1, the core pileup was large enough that all photons
within a $3''$\ of the source migrated to grade 7 and were
automatically rejected; nearby regions somewhat further off-axis were
also contaminated by pile-up.

Despite this, the source spectrum can be extracted from the ``transfer
streak,'' which appears as a line connecting the detector aimpoint and
the chip readout in Figure~\ref{fig:image}.  During the 42 ms frame
transfer of the ACIS CCDs, X-rays from the source continue to arrive
at the aimpoint and are as a result ``mis-positioned'' along the axis
of the transfer by the \chandra processing software.  In 9.74 ksec of
observing, the ACIS spends $\sim 128$\,seconds in frame transfer mode. 
We extracted $\sim 31,000$\ events in an $11\times908$\ pixel strip
around this streak (avoiding the near-aimpoint region which will be
affected to a lesser degree by pileup).  We were also concerned about
pileup in the transfer streak.  However, the streak has only
$10.3\pm3.2$\,counts per 3.2 second CCD frame spread over a 11x908
pixel region, so pileup there is not significant.
  
The in-flight spectral response of the ACIS-I detectors during frame
transfer is not yet specifically calibrated.  We assumed all the
events in the transfer streak originated on-axis with the normal
effective area in calculating the detector response, but there is a
known effect that slightly reduces the quantum efficiency of the CCDs
for photons above 3-4 keV (M. Bautz 2002 private communication).  We
also found that the ratio of grade 0 (single-pixel) events to all good
events (grades 0,2,3,4, and 6) in the transfer streak was
significantly lower than elsewhere on the ACIS-I chips.  This may or
may not affect the spectrum; the ACIS calibration team is working on
the problem.

Other non-contemporaneous X-ray observations of GX~13+1 exist.
GX~13+1 is regularly observed by the RXTE All-sky Monitor (ASM),
although unfortunately it was not observed for a period of $\sim
10$\,days surrounding the Chandra observation.  The long-term ASM
lightcurve shows that the luminosity in the 1.3-12.1 keV bandpass is
relatively constant, with intrinsic variations of $\sim 50\%$\,(Homan
\etal 1998).  GX~13+1 was also observed by ASCA for 18.8 ksec in
September 1994.  We extracted the GIS data, which have nearly
$10^6$\,counts; the source was blocked out in the SIS detectors.  The
flux observed by ASCA in the 1-10 keV band is $\sim 25$\% larger than
was observed by \chandra, likely due to fluctuations in the source
itself.

The observed halo intensity at energy $E$\ is proportional to the
observed source flux at energy $E$\ and the column density of dust
along the line of sight.  To estimate these we fit a range of spectral
models and examined the results.  Intrinsically, the X-ray spectra of
LMXBs vary widely due to the many possible viewing angles and emission
mechanisms.  We therefore considered a number of different models,
following the Christian \& Swank (1997) analysis of the X-ray emission
for a sample of LMXBs.  We used blackbody (BB), power law (PL), and
thermal bremsstrahlung (TB) models, along with a disk blackbody (DBB;
see Mitsuda \etal 1984, Makishima \etal 1986) model.  In addition, we
fit the spectrum using a series of top-hat functions, roughly matching
the ACIS-I resolution by using 100 eV spacing from 1-2 keV and 200 eV
spacing from 2-4 keV.  This approach measures the observed flux (with
errors) in each 100 or 200 eV band without reference to any specific
model.

\begin{figure}
\includegraphics[totalheight=3in]{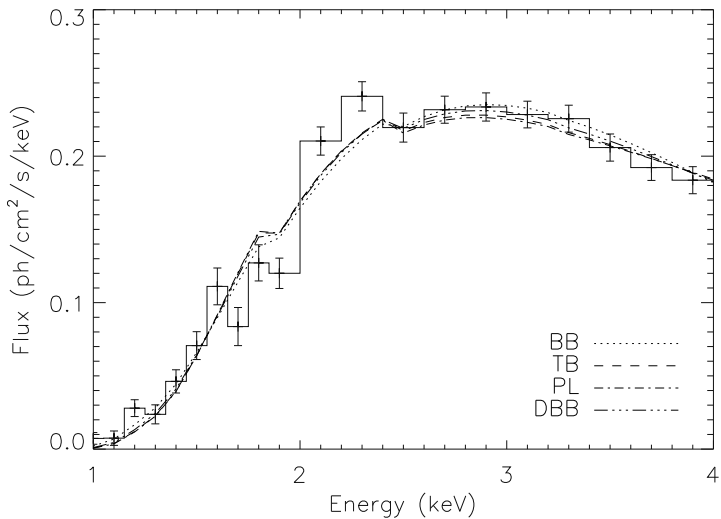}
\caption{Best-fit X-ray flux from GX~13+1 after fitting the observed
spectrum with blackbody (BB), power law (PL), thermal bremsstrahlung
(TB), and a disk blackbody (DBB) model.  The histogram shows the
fluxes and errors found using a model consisting of a sum of top-hat
functions.  Note that despite the wide range of models, the flux in
each energy band varies only slightly.\label{fig:TotFlux}}
\end{figure}

All of these models led to a formally acceptable value of $\chi^2$,
due to the low count rate and energy resolution in the transfer streak
data.  We calculated the X-ray flux as a function of energy for each
of these models, as shown in Figure~\ref{fig:TotFlux}.  The summed
top-hat model results in similar fluxes as the different spectral
models, and the errors on the top-hat amplitudes are on the same scale
as the variations in the spectral models.  We will use the top-hat
fluxes and errors for the following analysis.

The best-fit column density for these models varied from 1.99 to
4.17$\times10^{22}$\,cm$^{-2}$\ depending on the source model used.
Given the issues with the transfer streak detector response, we
decided not to use these \NH results.  The absorption column density
has been measured with other X-ray satellites; Ueda \etal (2001) found
\NH $= 2.9\pm0.1\times10^{22}$\,cm$^{-2}$\ from ASCA GIS and SIS
observations (with $> 10^6$\,counts), which agrees with the value
found by Schulz, Hasinger \& Tr\"umper (1989) from EXOSAT ME data.
Garcia \etal (1992) placed an upper limit of $A_V \lesssim 14.4$\,
based on measurements in the J, H, and K bands and assuming the
spectrum rose no faster than the Jeans tail of a blackbody.  Using the
conversion from \NH/$A_V = 1.9\times 10^{21}$\,cm$^{-2}$/mag (Seward
2000), this is equivalent to \NH $\lesssim 2.7\times10^{22}$\,cm$^{-2}$.

\subsection{Imaging Analysis}

Measuring the radial profile of GX~13+1 is a multi-step process.  We
began by removing any serendipitous sources from the image.  We then
selected an energy and a radial grid, based on the instrumental
response and the quantity of data available.  We filtered the data
using the CCD energy measurement, and extracted the counts in the
radial profile using concentric annuli centered on the source.  Then
we made exposure maps for each energy band.  After filtering out the
regions with serendipitous sources, we extracted the effective
``radial exposure'' using the same concentric annuli, for each energy
band.  Dividing the radial profile of the counts in each band by the
effective area and exposure time gave us the corrected radial profile
in units of photons cm$^{-2}$ s$^{-1}$ arcmin$^{-2}$\ per energy band.  We
then divided this by the observed source flux in each band to get the
fractional radial profile in units of source fraction arcmin$^{-2}$.
This result is really the X-ray halo convolved with the \chandra PSF,
plus the PSF from the source itself.  However, the \chandra PSF is
small enough that we will neglect any scattering of the X-ray halo by
the telescope.

We searched for sources in the field of view using the CIAO tool {\it
celldetect}.  Since removing a false source merely reduces the signal
to noise slightly, while including a source mistakenly would add a
systematic error, we decided to accept any source found with
significance greater than $4\sigma$.  This resulted in six sources,
listed in Table~\ref{tab:srcs}.  A search using {\it Simbad}\ showed
no known sources with $10''$\ of these positions.  The {\it Simbad}\
search found 39 objects in the field of view, including stars, IRAS
objects, and a supernova remnant (G13.5+0.2).  None of these sources
could be detected in the data, although we note source D is inside the
remnant.

\begin{table}
\caption{Serendipitous Source List\label{tab:srcs}}
\begin{tabular}{llll}
\hline \hline Source & RA (J2000) & Dec (J2000) & Net Flux (cts/ksec)
\\ \hline  
B & 18:13:55.833 & -17:06:28.09 &  76.8958 \\
C & 18:14:37.346 & -17:10:39.70 &  4.77356 \\
D & 18:14:10.560 & -17:11:17.01 &  2.17923 \\
E & 18:15:04.775 & -17:04:46.79 &  3.00942 \\
F & 18:14:02.059 & -17:01:33.98 &  4.25470 \\
G & 18:14:06.447 & -17:03:00.61 &  7.05657 \\ \hline
\end{tabular}
\end{table} 

The X-ray halo of GX~13+1 fills the ACIS-I field of view, as can be
seen in Figure~\ref{fig:image}.  However, the central $3''$\ region
around the source has no data due to severe pileup, so the source
position must be inferred.  We estimated the source position by
assuming it was on the line defined by the transfer streak and
centering using the contours of the data.  We found the position of
GX~13+1 to be RA,Dec = 18:14:31.065, -17:09:26.02 (J2000), with an
estimated error of $0.3''$.  This is $0.3''$ distant from
the position measured by Berendsen \etal (2000) using radio
observations of GX~13+1 with the ATCA.  The halo appears to be
radially symmetric about the source, as expected from previous
observations.  We confirmed this by examining the data in various
radial bins as a function of azimuthal angle.  The data showed no
significant variation as a function of angle.  We were therefore
justified in extracting the radial profile of the X-ray halo by
summing the data in annular rings in our selected energy bands.

\begin{figure}
\includegraphics[totalheight=2.2in]{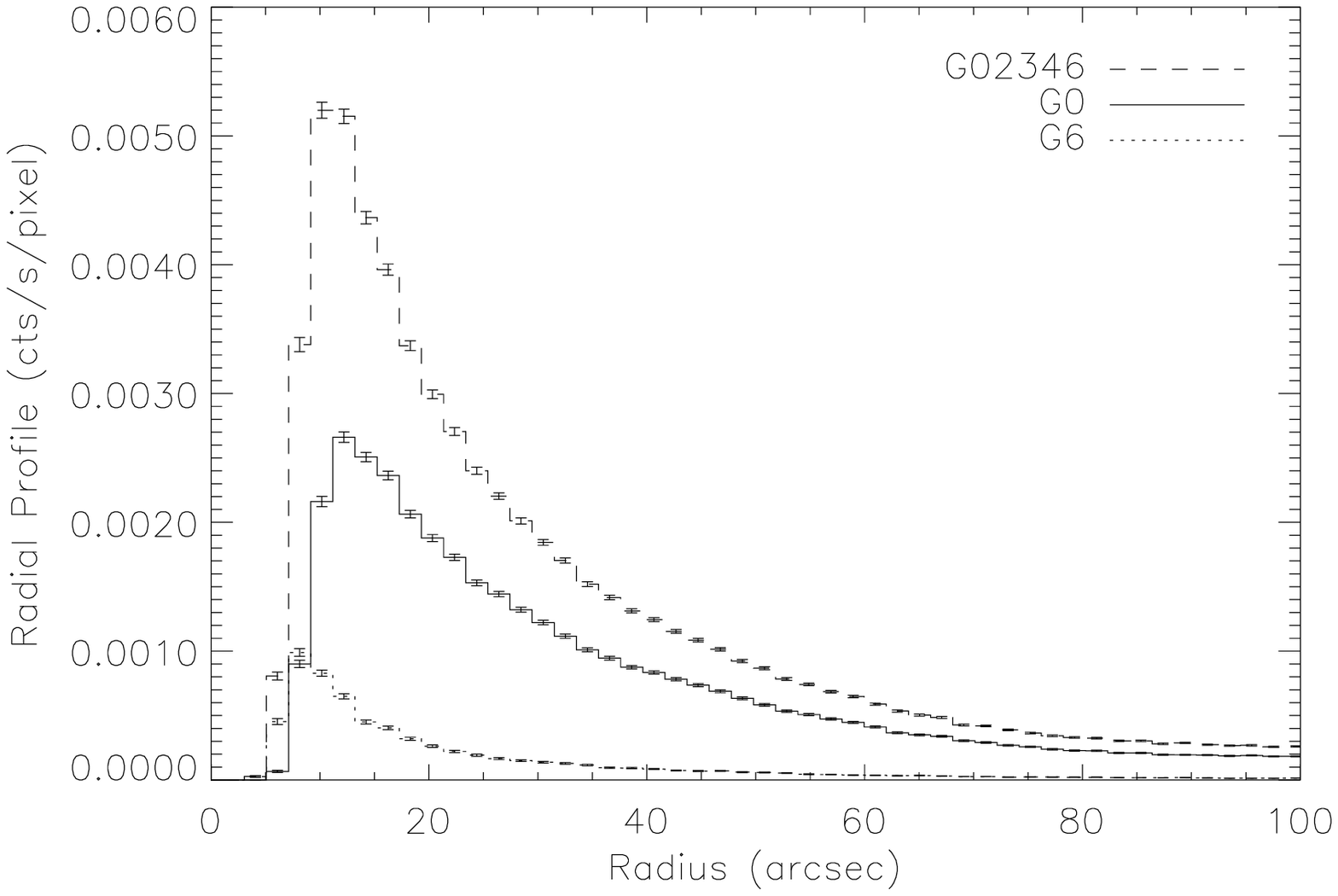}
\includegraphics[totalheight=2.2in]{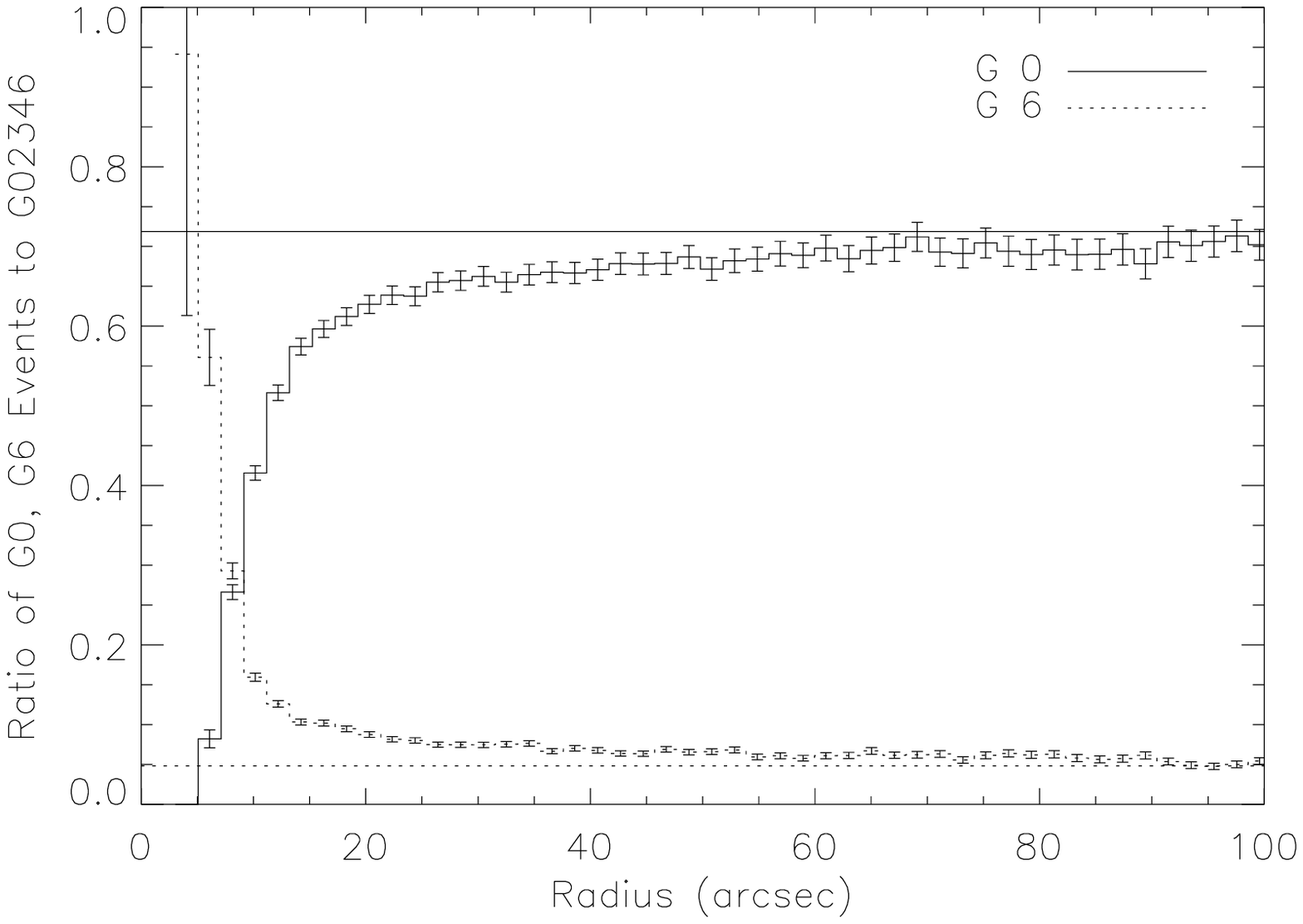}
\caption{(a) The surface brightness of GX~13+1 as a function of
radius, including both grade 0, grade 6, and the sum of all ``good''
grades 0, 2, 3, 4, and 6.  Near the source, marginal grade 6 events
dominate, while far from the source single-pixel grade 0 events do.
(b) The ratio of grade 0 and grade 6 to the sum of all good events, as
a function of radius.  The horizontal line shows the expected value
for grade 0 events far from the source.  Within $50''$\ of the source,
pileup is a noticeable issue.\label{fig:radial_grade}}
\end{figure}

Pileup substantially affects the central radial bins, so we had to
determine where it drops off.  This is difficult to estimate {\it a
priori}, since pileup can occur not only when two X-rays strike the
same pixel within a single frame time, but also when the two X-rays
hit adjacent pixels.  Depending on the details of the interaction of
the X-ray in the CCD, pileup may lead to events not being recognized,
or being rejected, or being identified with the wrong energy and/or a
modified grade.  We therefore estimated the pileup rate
phenomenologically by examining the radial profiles of different grade
events.  When the count rate is low, the ratio of event grades should
tend to a constant which depends on the spectrum.  In
Figure~\ref{fig:radial_grade}(a)\ we plot the radial profile of good
(grade 0,2,3,4, and 6) events, along with the profile of just grade 0
(single-pixel), and grade 6 (four-pixel) events.
Figure~\ref{fig:radial_grade}(b) shows the ratio of grade 0 and grade
6 events to the total as a function of radius, along with the limiting
ratio found far from the source.  This is not a perfect estimate of
the ratio close to the source since the X-ray halo will have some
spectral differences from the source, but we do not expect the
difference to be substantial.  As Figure~\ref{fig:radial_grade}(b)
shows, pileup/grade migration is a noticeable effect as far as
$50''$\ from the source.  We can derive a similar result analytically
as well.  According to the \chandra proposers' guide, a pileup
fraction of 10\% will impact the resulting image or spectrum.  Using
the 3.2 second ACIS frame time and a poisson-distributed count rate,
10\% pileup (or two photons in a single $3\times3$\ pixel grid in one
frame) corresponds to a rate of 0.0037 cts\,s$^{-1}$\,pixel$^{-1}$.
As Figure~\ref{fig:radial_grade}\ shows, this is the rate found within
$\sim 20''$\ of the source, where the effect of pileup is most
severe.  To avoid contamination by pileup, we limited our halo fits to
angles beyond $50''$. 

Before extracting the radial profile of GX~13+1, we had to select the
energy and radial binning.  To limit complications due to multiple
scattering and problems with the RG approximation, we decided to use
only the X-rays above 2 keV.  We divided the dataset into 10 energy
bands, spaced every 0.2 keV from 2.0-4.0 keV in order to match the
ACIS-I resolution.  To effectively use \chandra's angular resolution
while also roughly equalizing the number of counts in each annulus, we
used 99 log-spaced annuli from $10''$\ to $600''$\ (plus one from
0-$10''$).  After extracting the radial profiles of the data, we
calculated the exposure maps for each energy band and used the same
annuli to extract the total effective area at each energy and radius.
The serendipitous sources were removed both from the data and the
exposure maps, so the results are self-consistent.  We then divided
the radial profile of event data by that of the exposure map to get a
result in physical units of photons
cm$^{-2}$\,s$^{-1}$\,arcmin$^{-2}$.  This approach does not take into
account the varying response of the CCDs.  On-axis at 1.5 (4.5) keV,
the FWHM is $\sim 160$ ($\sim 300$) eV, while at the chip readouts, it
is $\sim 80$ ($\sim 130$) eV.  Due to the physical layout of the
ACIS-I CCDs, small off-axis angles are all far from the chip readouts,
while large off-axis angles will contain points both near and far from
the chip readouts (Chandra Proposers' Observatory Guide, Rev. 4.0,
pp. 74, 86).  As a result, there will be a systematic effect in the
``average'' ACIS response as a function of off-axis angle.  Using
energies between 2-4 keV summed in broad 200 eV bands will minimize
these effects.

The final stage of the analysis is to remove the \chandra PSF, which
was extremely complicated.  The CIAO tool {\it mkpsf}\ only models the
core of the PSF and is not useful here since we are interested in the
PSF $1'$\ or more from the source.  With the help of the CXC
calibration team, we ran the \chandra raytrace model ``SAOsac,''
(release as of 9/18/01) at a number of energies.  SAOsac matches the
observed core of the PSF well, but underestimates the wings of the PSF
(T. Gaetz, 2002, private communication).  The exact amount of the
disagreement depends on the X-ray energy and the angle from the
source; $500''$\ from the source it varies from a factor of 2 at 1.5
keV to a factor of 6 at 4.5 keV, comparing the pre-flight calibration
data to the model\footnote{Chapter 15 of the XRCF report,
http://hea-www.harvard.edu/MST/simul/xrcf/report/index.html}.

Unfortunately, the systematic underestimate of the scattering power
beyond $50''$\ led to serious problems in interpretation.  To address
this, the \chandra calibration team observed the bright,
lightly-absorbed X-ray binary Her X-1 for 50 ksec on July 1, 2002.
Her X-1 is lightly reddened, with $E(B-V) = 0.05$\,magnitudes (Liu,
van Paradijs, \& van den Heuvel 2001), which corresponds to a column
density \NH $\approx 3\times10^{20}$\,cm$^{-2}$, although some of this
may be internal.  The total Galactic column density in the direction
of Her X-1 is \NH $= 1.8\times10^{20}$\,cm$^{-2}$\ (Dickey \& Lockman
1990).  As a result, the dust-scattered halo is more than $100\times$\
weaker than in GX~13+1, and the telescope PSF dominates the radial
profile.  This Her X-1 image, which clearly shows vignetting from the
mirror support structures, confirms this; in GX~13+1, these features
are blurred by the X-ray halo.

\begin{figure}
\includegraphics[totalheight=3in]{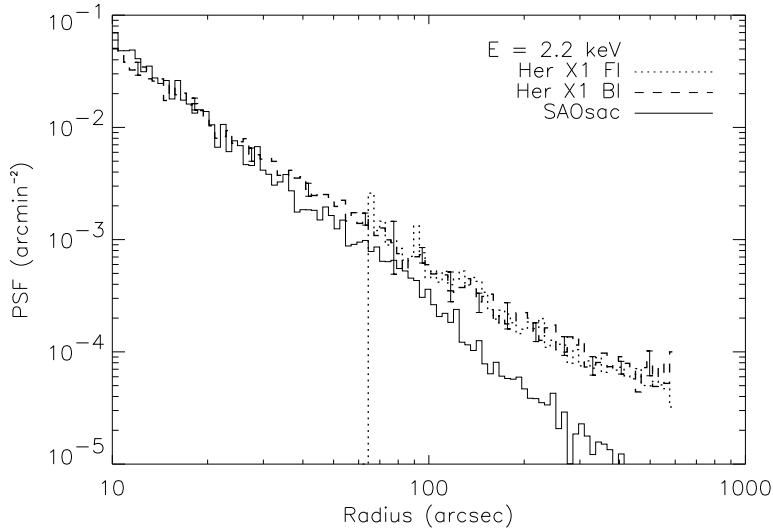}
\caption{The radial profile of Her X-1 in the FI and BI chips between
2.1-2.3 keV, along with the modeled PSF from SAOsac.  Her X-1 was on
the BI chip S3, so the FI data is available only beyond $\approx
65''$\ off-axis.  Her X-1 is very lightly absorbed; based on the
measured spectrum and assuming a smooth distribution of MRN dust, we
calculated that the total X-ray halo will have less than 1.5\% of the
direct photons.\label{fig:PSFcompare}}
\end{figure}
The Her X-1 data are being analyzed in detail by the CXC calibration
team.  We did a preliminary analysis using the same procedure as used
for the GX~13+1 data, extracting the spectrum from the transfer streak
and fitting the flux in 100 eV and 200 eV bands.  The observation was
done with Her X-1 on ACIS-S3 (a back-illuminated (BI) chip), along
with the front-illuminated (FI) I2, I3, and S2 chips.  We extracted
the radial profile for the BI and FI chips separately, and compared
them to each other and to the SAOsac PSFs.  The FI and BI profiles
agreed with each other at all off-axis angle, and with the SAOsac
values for off-axis angles less than $50''$\ (see
Figure~\ref{fig:PSFcompare}).  As expected however, beyond $50''$\
off-axis, the SAOsac PSF underestimated the data by factors of 2-10,
depending on the angle and the X-ray energy.  Background appeared to
be negligible since the FI and BI profiles match each other even at
large angles, despite the significant differences in the background
rates on the two types of chips.  We summed the FI and BI datasets to
get the best measurement of the PSF at each energy.

\section{Results}

\begin{figure}
\includegraphics[totalheight=3in]{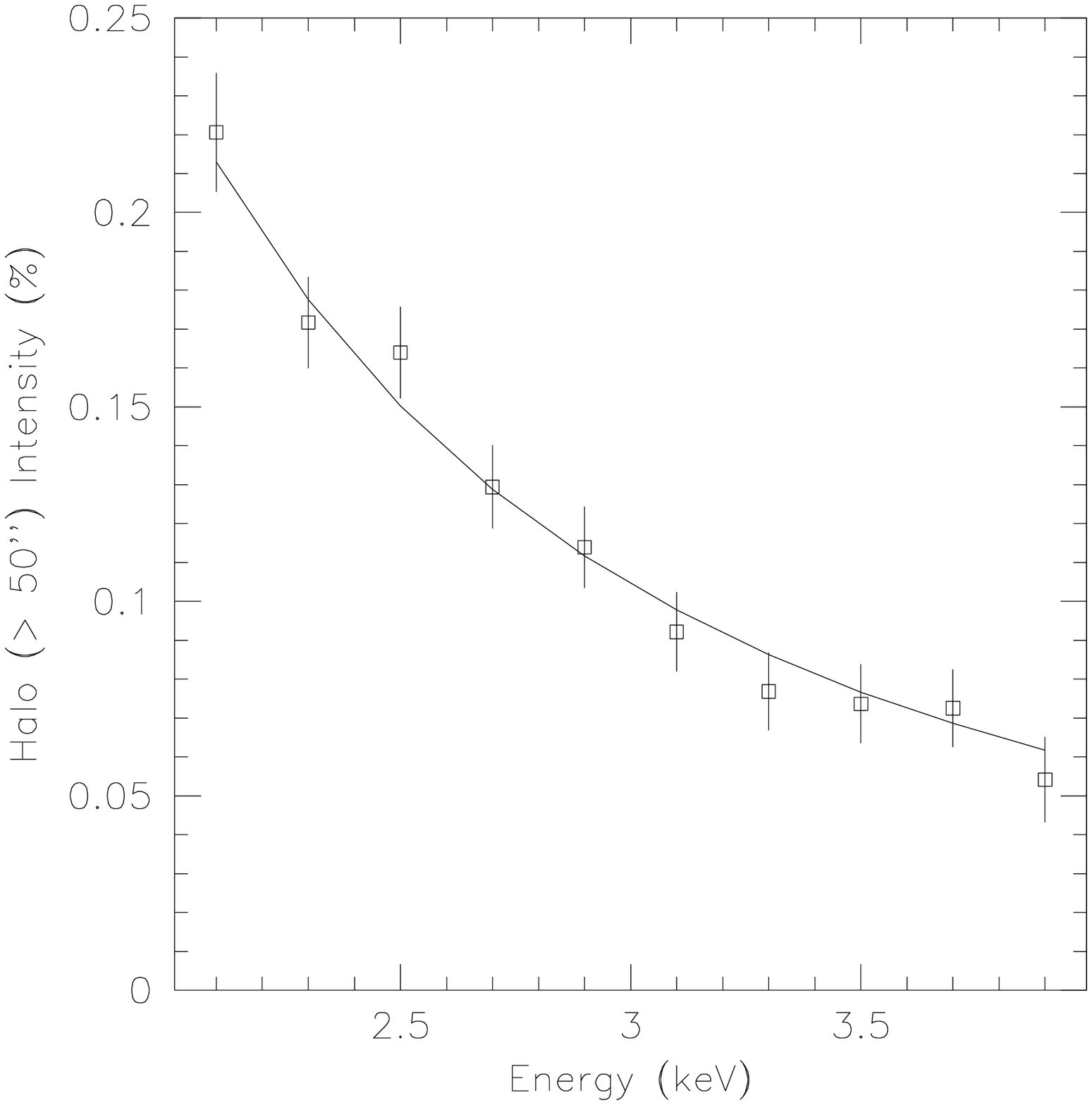}
\caption{The fractional halo intensity (relative to the source flux),
as a function of energy along with the best-fit curve $I(E) =
0.939\pm0.028 E_{keV}^{-2}$. \label{fig:HaloIntensity}}
\end{figure}

We have now measured the radial profile, in units of fraction of
source flux per square arcminute, for both GX~13+1 and Her X-1.  Since
the source flux has been factored out, subtracting the Her X-1 profile
(the best-available point source) from the GX~13+1 profile gives the
halo's surface brightness relative to the source flux as a function of
energy and off-axis angle, or the ``halo fraction'' $H_f(E, \theta)$.
We can then calculate the total observed X-ray halo fraction as a
function of energy:
\begin{equation}
I(E) = \int_{50''}^{600''} H_f(\theta, E, \Delta E)\times 2 \pi \theta\,
d\theta
\label{eq:HF}
\end{equation}
which is also shown for GX~13+1 in Figure~\ref{fig:HaloIntensity}.
Figure~\ref{fig:HaloIntensity}, which strongly increases at low
energies, shows why ROSAT (with a 0.08-2.9 keV bandpass) measured a
larger total halo fraction (34\%) for GX~13+1 than either of the
Einstein (with a 0.1-6 keV bandpass) measurements of 17-18\%.  

Using the RG approximation, this result be directly compared to the
observable dust grain parameters.  As discussed in Mauche \& Gorenstein
(1986), the total scattering due to dust grains is simply $\int da\,
n_{gr}(a) \sigma(a) D$, where $n_{gr}(a)$\ is the number density of grains
of size $a$, $D$\ is the source distance, and $\sigma(a)$\ is the
total scattering cross section, which can be found by integrating
Equation~(\ref{eq:RG})\ over all angles.  The final result is
\begin{equation}
I(E) \approx 0.2 \Big({{2Z}\over{M}}\Big)^2
\Big({{\rho}\over{3\hbox{g cm}^{-3}}}\Big)^2 
\Big[{{F(E)}\over{Z}} \Big]^2
D_{\hbox{kpc}} E_{\hbox{keV}}^{-2} 
\int da\,n(a)\,\Big({{a}\over{0.1\mu\hbox{m}}} \Big)^4 
\Big({{n_{gr}(a)}\over{10^{-12}\hbox{cm}^{-3}}}\Big) 
\label{eq:RGint}
\end{equation}
We can approximate some terms in Equation~(\ref{eq:RGint}), as $2Z/M
\approx 1$\ and for $E > 2$\,keV, $F(E) \approx Z$.  With
some manipulation, we then get:
\begin{equation}
I(E) = 80.7 \Big({{\NH}\over{10^{22}\hbox{cm}^{-2}}}\Big)  
\Big({{\rho}\over{3\,\hbox{g cm}^{-3}}}\Big) E_{\hbox{keV}}^{-2}
\int da\,n(a) \Big({{a}\over{0.1\mu\hbox{m}}} \Big)
{{m_{{gr}}(a) }\over{m_{\rm H}}} 
\label{eq:HaloTest}
\end{equation}
where $m_{gr}(a)$\ is the mass of a grain of radius $a$.  Any dust
grain model must include the dust to gas mass ratio ($m_{gr}/m_H$),
the grain density $\rho$, and the size distribution of the grains
$n(a)$.  Therefore, given a dust model and a column density \NH we can
calculate the coefficient term in Equation~(\ref{eq:HaloTest})
directly.  Since we used the total cross section for scattering, this
result is independent of the position of the dust along the line of
sight.

We can measure the coefficient term directly from the results we found
using Equation~(\ref{eq:HF}) by fitting a function with the $E^{-2}$\
energy dependence of Equation~(\ref{eq:RGint}).  We found, for
GX~13+1,
\begin{equation}
I(E) = H_f E_{\hbox{keV}}^{-2} = 0.939\pm0.028 E_{\hbox{keV}}^{-2}
\end{equation}
which is also plotted in Figure~\ref{fig:HaloIntensity}; the error
purely statistical, and does not include any systematic errors.  This
halo coefficient $H_f$\ is an underestimate, since we only
include scattering for $50'' < \theta < 600''$.  We experimented by
changing the upper and lower limits, and found reducing the upper
limit by 50\% to $400''$\ changed $H_f$\ only slightly, to
$0.89\pm0.02$.  However, increasing the lower limit by 50\% to $75''$\
led to $H_f = 0.73\pm0.02$, a more significant change.  We therefore
tried reducing the lower limit, although this increases systematic
errors due to pileup.  A lower limit of $20''$\ led to $H_f =
1.28\pm0.04$, and a lower limit of $10''$\ gave $1.41\pm0.04$; within
$10''$\ of the source the data are visibly affected by pileup.  We did
a linear extrapolation to the lower limit of $0''$\ based on these
results and found $H_f = 1.5^{+0.5}_{-0.1}$.  These errors are not
statistical, but rather are estimated from the variation
with the lower limit on the angle.  We used a large positive error
term because we must assume that the scattered surface brightness
within $10''$, which we cannot measure directly, does not increase
significantly.

Since the X-ray halo is a function of the size, position, and
composition of the dust grains, as well as the source flux and
absorption column, to go further in the analysis we must make some
assumptions about the dust grains.  We first consider the
smoothly-distributed dust model, where the dust grain number density
is constant along the line of sight.  We compare three dust grain
models: the MRN model, the Weingartner \& Draine (2001; WD) model, and
the ``extended'' MRN model (hereafter the XMRN model) described by
Landgraf \etal (2000) and used by Witt, Smith, \& Dwek (2001).  This
model is based on the {\it in situ}\ measurements made by {\it
Ulysses}\ and {\it Galileo}\ in the heliosphere, and has the same
total mass as the MRN model, but extends the MRN size distribution to
$2.0\,\mu$m, with a break at $0.5\,\mu$m from a $-3.5$\ power law to
$-4.0$\ for larger grains.

\begin{figure}
\includegraphics[totalheight=2in]{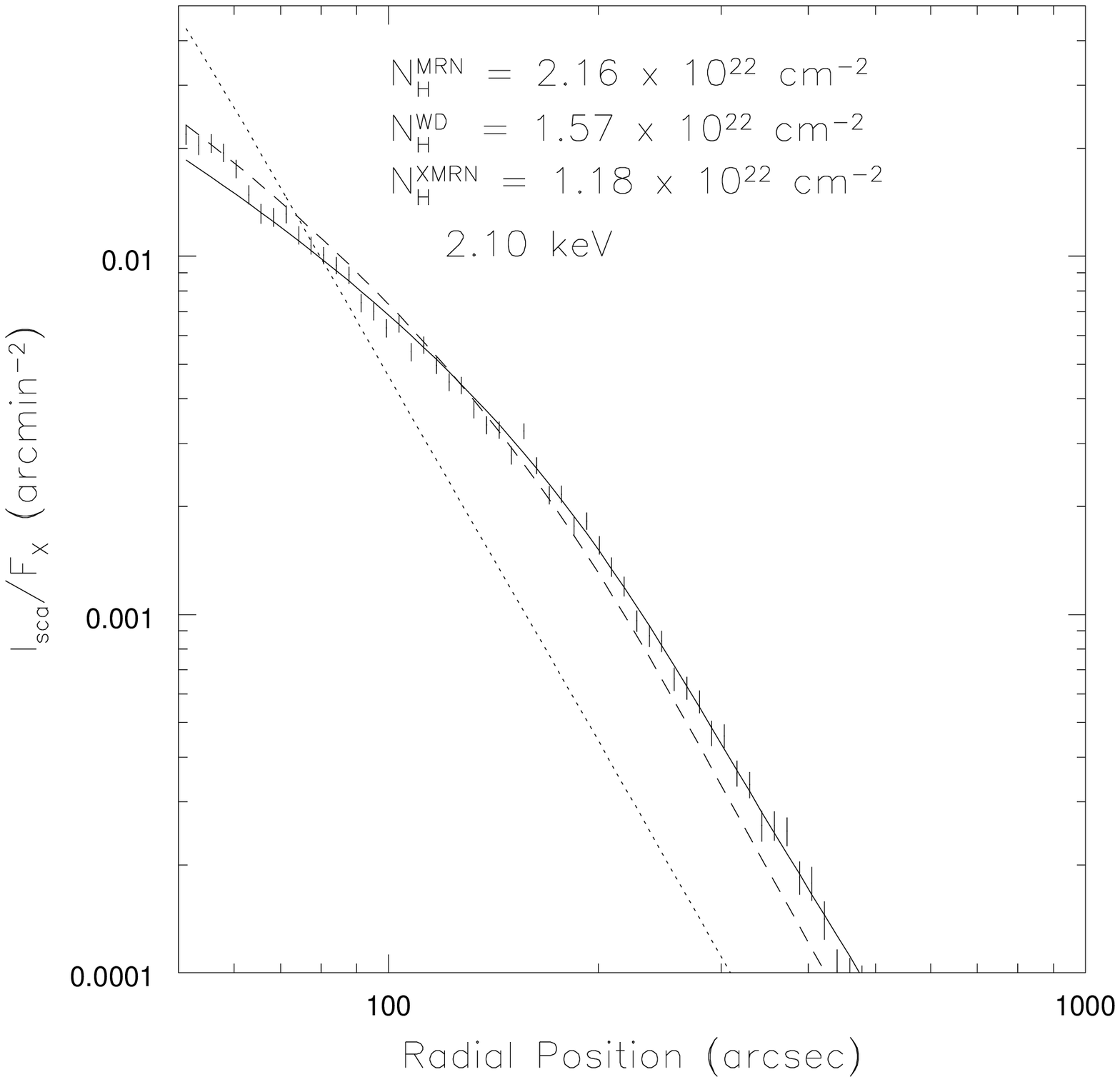}
\includegraphics[totalheight=2in]{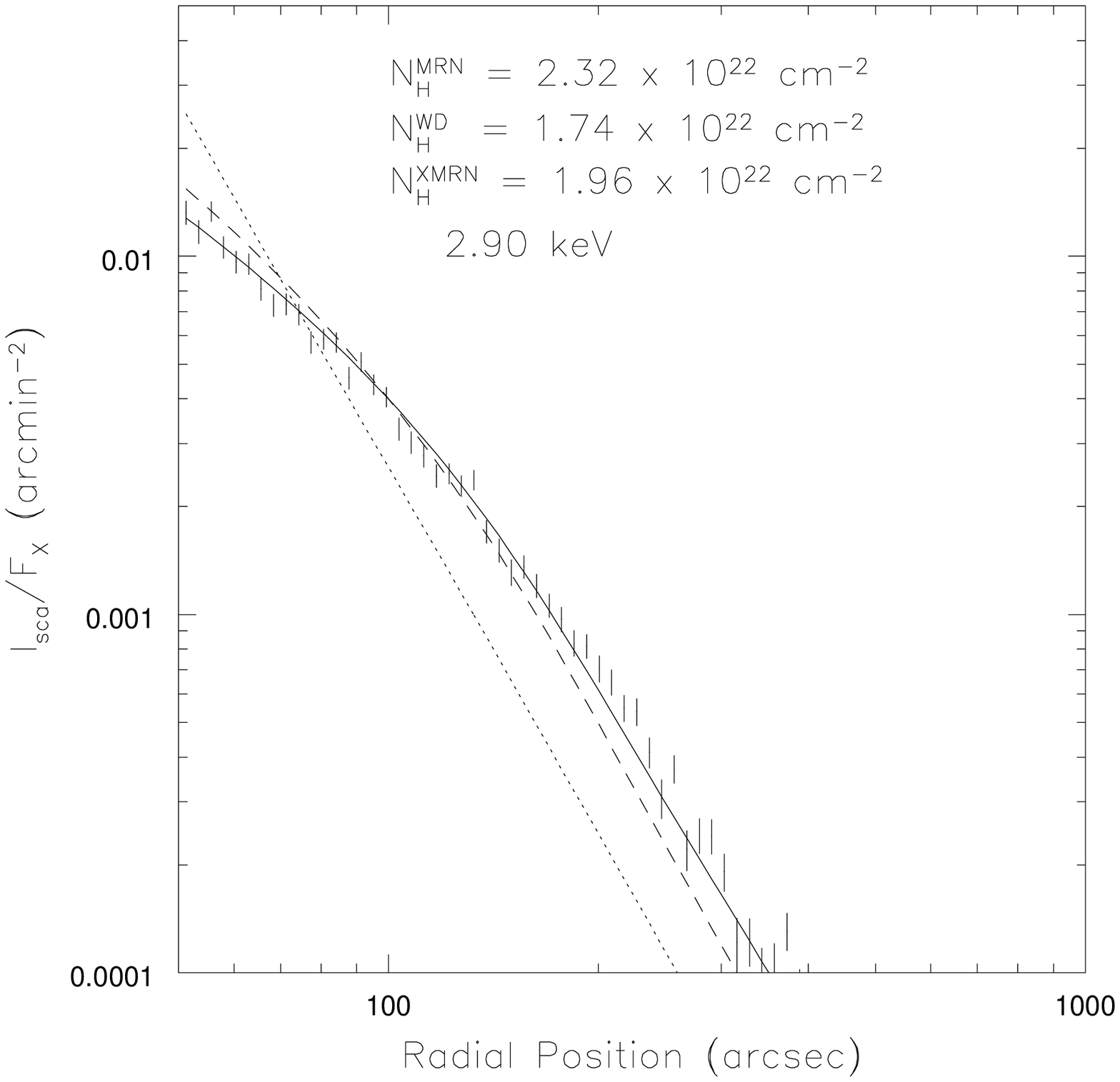}
\includegraphics[totalheight=2in]{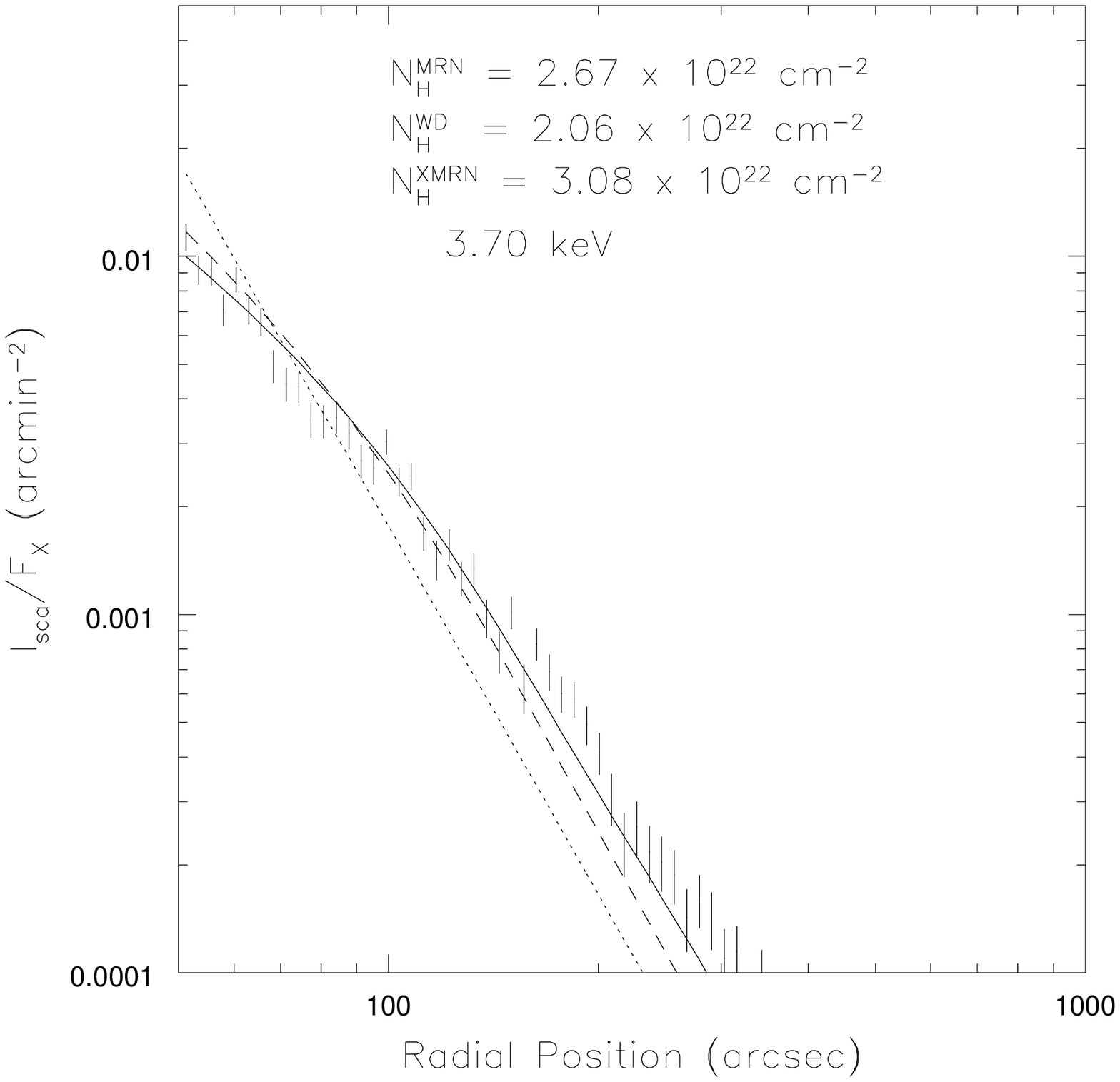}
\caption{X-ray halos observed at $2.1\pm0.1$, $2.9\pm0.1$, and
$3.7\pm0.1$\,keV, fit using smoothly-distributed dust described by the
MRN (solid line), WD (dashed line), and the XMRN (dotted line) models.
The only free parameter \NH differs substantially between the three
grain models; for the XMRN model it appears to systematically increase
with X-ray energy.
\label{fig:SmoothModels}}
\end{figure}

In Figure~\ref{fig:SmoothModels}\ we show our best-fit results at
three energies, using the MRN, WD, and the XMRN dust grain models.
The only free parameter in these fits is the total dust grain column,
which we allowed to vary independently in each energy band.  A number
of features are immediately apparent in these results.  First, the
XMRN model is consistently poor, overestimating the halo at small
angles and underestimating it at large angles.  At 2.1 keV, this could
be partially due to the RG approximation breaking down, but RG should
be adequate at 3.7 keV where it is still clearly a poor fit.  Although
we cannot rule out models such as the XMRN, this dataset certainly
does not require them.  Secondly, we see that the MRN and WD models
both match the halo profile quite well, although the MRN model tends
to be the better fit.  Including all 10 energies from 2-4 keV, and
allowing the column density \NH to vary with energy, for the MRN model
we found $\chi^2_{red} = 1.6$, and for the WD model $\chi^2_{red} =
5.2$\ (with 590 degrees of freedom).  If we require the column density
to be constant, there are 599 degrees of freedom and the best fit for
the MRN model is \NH $= 2.22\pm0.01\times10^{22}$\,cm$^{-2}$\ with
$\chi^2_{red} = 2.1$.  For the WD model, it is \NH $=
1.65\pm0.01\times10^{22}$\,cm$^{-2}$\ with $\chi^2_{red} = 5.8$.

We can check this by comparing the measured halo coefficient of $H_f
\approx 1.5$\ to the values calculated using
Equation~(\ref{eq:HaloTest}) and \NH $= 2.9\times10^{22}$\,cm$^{-2}$.
For the MRN, WD, and XMRN models, we calculated halo coefficients of
$1.9$, $2.8$, and $13.1$\ respectively.  As expected, if we use the
best-fit \NH values for the MRN and WD models, we get much better
results: 1.5 and 1.6, respectively.  This also shows that the best-fit
column density found by fitting the X-ray halo using the MRN model
will generally be $\sim 45$\% larger than fits using the WD model.  

\begin{figure}
\includegraphics[totalheight=3.5in]{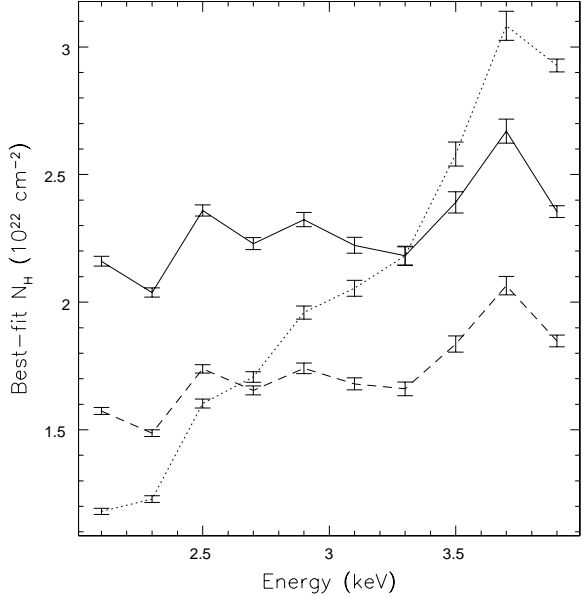}
\caption{The best-fit \NH as a function of X-ray energy both for the
MRN model (solid line), the WD model (dashed line), and the XMRN model
(dotted line).  The slight increase with energy, especially above 3.3
keV, is clear in all cases although it is far worse in the XMRN case.
The error bars are statistical.  The systematic error is dominated by
the flux measurement, and there may be calibration problems with the
higher energies. \label{fig:NHvsE}}
\end{figure}

The best-fit column density \NH shows a slight increase with the X-ray
energy for the MRN and WD models, and a significant one for the XMRN
model (see Figure~\ref{fig:NHvsE}).  The errors shown in
Figure~\ref{fig:NHvsE}\ are dominated by the errors in the flux
measurement.  Since we fix each energy band width at 200 eV and do not
consider the effect of the varying energy resolution of the ACIS
chips, these errors may be somewhat underestimated.  It is also
possible the quantum efficiency of transfer streak may be lower than
expected (M. Bautz, 2002, private communication).  If true, this would
mean we underestimated the true flux, which would lead us to a larger
\NH to compensate.  For the XMRN model, however, the energy dependence
is much steeper over the entire 2-4 keV energy range.  The source of
the problem could be our use of the RG model, or or possibly it is
just an artifact caused by the poor overall fit of the XMRN model.  We
expect to investigate this in a future paper, including applying a
full Mie solution to the XMRN model.

\subsection{Molecular clouds}

\begin{figure}
\includegraphics[totalheight=2in]{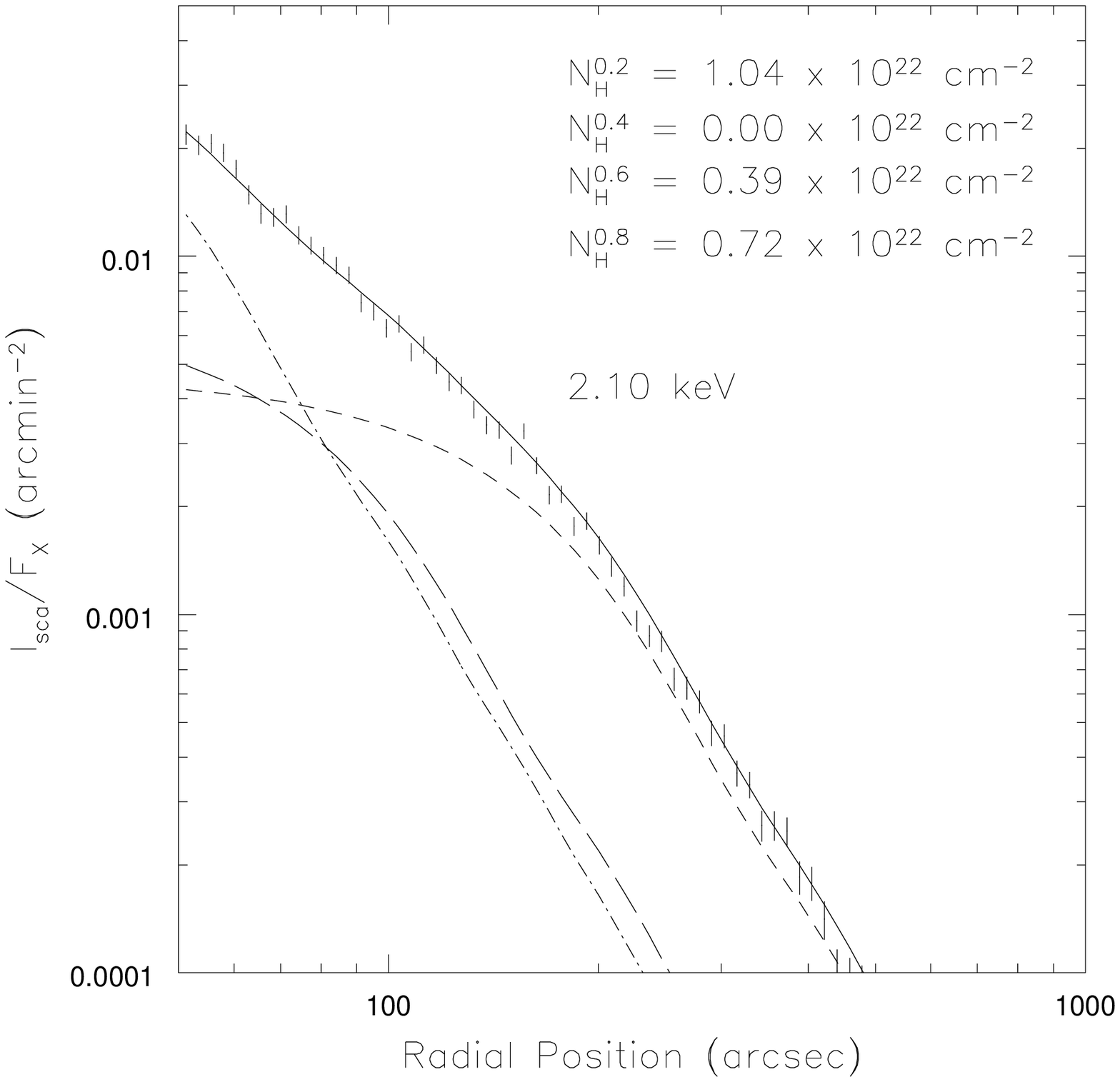}
\includegraphics[totalheight=2in]{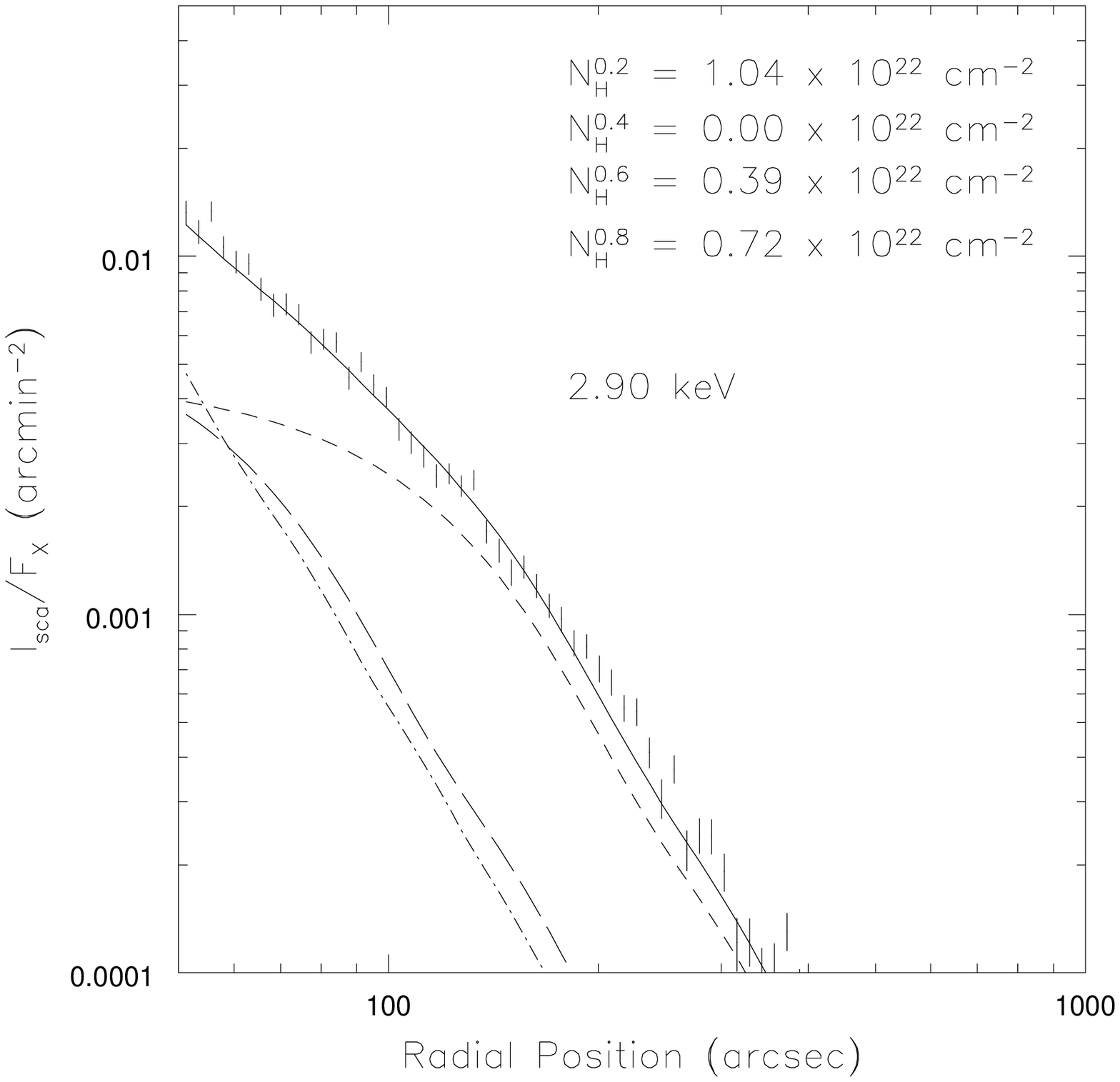}
\includegraphics[totalheight=2in]{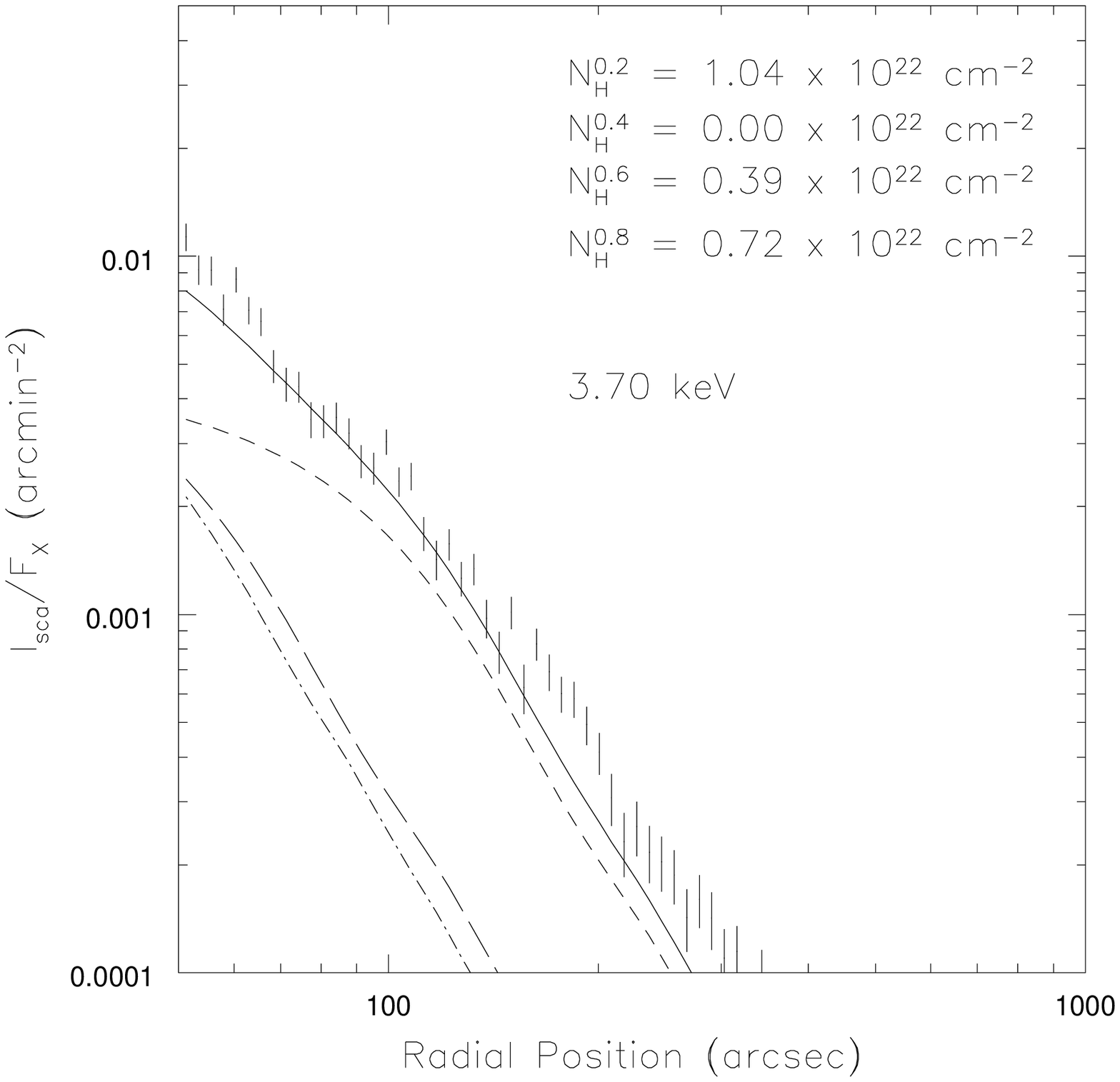}
\caption{X-ray halos observed at $2.1\pm0.1$, $2.9\pm0.1$, and
$3.7\pm0.1$\,keV, fit using four evenly-spaced clouds of dust (using
an MRN model) at 20\%, 40\%, 60\%, and 80\% of the distance to
GX~13+1.  Each dust cloud has a variable amount of dust parameterized
by \NH.  The best-fit total scattering is given by the solid line.
The scattering from the cloud at 20\% is shown by the dashed line, at
60\% by the long-dashed line, and at 80\% by the dot-dashed line.
\label{fig:ClumpModels}}
\end{figure}

We have assumed that the dust grains were smoothly distributed along
the line of sight ({\it i.e.}\ $f(x)\equiv1$).  We can relax this
restriction and assume instead that dust grains are found in clumps
along the line of sight.  This is probably a more realistic model, as
the sightline to GX~13+1 crosses two or three spiral arms (Caswell \&
Haynes 1987; Taylor \& Cordes 1993), spaced at about 20\%, 40\%, and
60\% of the distance to GX~13+1.  In Figure~\ref{fig:ClumpModels}\ we
show the profiles for E=2.1, 2.9, and 3.7 keV along with fits using
MRN dust and four ``clouds'' placed 20\%, 40\%, 60\%, and 80\% of
the distance to GX~13+1.  The amount of dust in each cloud was allowed
to vary independently, but was fixed to be the same value for each
X-ray energy.
These fits show one result similar to the smoothly-distributed dust
case: as the energy increases, the model tends to underpredict the
total halo.  Since the amount of dust in each cloud was not allowed to
vary with energy, the best-fit result ($\chi^2_{red} = 2.0$) slightly
exceeds the data at low energies and underestimates at high energies.
In addition, the total amount of dust used (\NH $=
2.15\times10^{22}$\,cm$^{-2}$) is similar to the value for the
smoothly distributed MRN model.

We also experimented with models using one or two clouds with variable
locations and column densities.  We could not find an adequate fit
with a single cloud model.  The lowest $\chi^2_{red} = 5.2$\ for the
MRN model had a cloud at $x=0.39$\ and column density \NH
$=1.65\times10^{22}$\,cm$^{-2}$.  Fitting two clouds with variable
positions and column densities, however, we found a minimum
$\chi^2_{red} = 2.31$, nearly as good a fit as Galactic arm model with
four fixed clouds.  The best-fit cloud parameters are \NH = $1.15 \pm
0.01 \times10^{22}$\,cm$^{-2}$\ at $x=0.204\pm0.007$\ and \NH =
$1.19_{-0.45}^{+0.02}\times10^{22}$\,cm$^{-2}$\ at $x=0.80\pm0.08$.

These fits show that two or three ``dust clouds'' can fit the data
nearly as well as the smoothly-distributed case.  One cloud must be
near GX~13+1 to reproduce the scattering seen at $\theta < 100''$\ and
the second cloud must be near the observer to cause the more extended
scattering.  In the four fixed cloud model, the clouds 40\% and 60\%
of the distance to GX~13+1 have the smallest column densities; in the
two cloud model, the data tend to the 20\% and 80\% positions
naturally.  This suggests that the nearest spiral arm contains
substantial dust, along with material near the Galactic center.

\section{Conclusions}

The high angular resolution of the \chandra telescope provides the
best data to date of any X-ray halo, while the energy range and
resolution allow us to explore the halo's behavior as a function of
energy with moderate resolution.  Although we limited our
investigation to off-axis angles greater than $50''$\ to avoid any
possible contamination by pileup, our models tended to fit reasonably
well for smaller angles and we may be able to relax this in the
future.  We also limited our analysis to energies higher than 2 keV,
to allow the use of the RG model and to avoid the difficulties
involved in including multiple X-ray scattering.  These issues can be
addressed with improved modeling techniques, which we plan to include
in a future paper.

We were able to measure, for the first time, the halo coefficient $H_f
= 1.5^{+0.5}_{-0.1}$\ by fitting the total halo intensity (relative to
the source intensity) as a function of energy to the equation $I(E) = H_f
E_{\hbox{keV}}^{-2}$.  Physically, $H_f$\ is the halo fraction at 1
keV, but using the RG approximation it can be compared directly to
dust model parameters via a simple calculation (see
Eq. (\ref{eq:HaloTest})).  The large errors in our measurement are due
largely to pileup; more X-ray halo measurements are clearly needed to
get more halo coefficients.  We expect this will soon provide a tight
constraint for dust models for the ISM.

We have found that models with low density ``fluffy'' dust are not
supported by these data.  The X-ray halo surface brightness is
proportional to $\NH \times \rho_{dust}^2$.  For the MRN and WD
models, we found \NH $=2.22\pm0.01\times10^{22}$\,cm$^{-2}$\ and
$1.65\pm0.01\times10^{22}$\,cm$^{-2}$, respectively.  The true column
density to GX~13+1 is difficult to measure, but we infer from infrared
observations that \NH $\lesssim 2.7\times10^{22}$\,cm$^{-2}$\ and from
the ASCA data that \NH $=2.9\pm0.1\times10^{22}$\,cm$^{-2}$.  So at
most, the true column density to GX~13+1 is only twice the lowest
value determined from the halo observations, which allows for a dust
grain density at most 30\% lower than the nominal value, and likely
less.  A ``fluffy'' dust model may exist that matches the
observations, but it would have to include substantially more dust
grains to maintain the total scattering cross section.  For example,
an MRN-type model with $\rho = \rho_{\rm MRN}/5$\ would need 25 times
more dust grains to achieve the same total scattering.

At the same time, the large dust grain model used in Witt, Smith \&
Dwek (2001) is also not supported by these data, in agreement with the
results of Predehl \& Schmitt (1995).  The large dust grains cause the
scattering at very small observed angles ($< 100''$) to increase
dramatically.  The ROSAT PSPC data used by Witt, Smith \& Dwek (2001)
had only one point at $100''$, which drove the fit to prefer the large
dust grain model.  {\it Chandra}'s superior angular and energy
resolution (and range) show that for this line of sight at least,
large dust grains are not significant.  Finally, we also found that
the data are better fit by the MRN model than the WD model, although
enough potential calibration issues remain to keep this an open issue.

We would like to thank Terry Gaetz for obtaining the Her X-1
calibration observations, Diab Jerius for helping us with SAOsac, and
the entire CXC calibration team for their substantial contribution to
this project.  We also thank Eli Dwek for many helpful discussions,
and the referee Peter Predehl for his comments and suggestions which
greatly improved this paper.  This work was supported by the Chandra
X-ray Science Center (NASA contract NAS8-39073) and NASA Chandra
observation grant GO0-1107X.

\end{document}